\documentclass[showpacs, preprintnumbers, nofootinbib, aps, prd,
            superscriptaddress,10pt, showkeys, notitlepage]{revtex4-1}

\usepackage{graphicx, amssymb, amsmath, amsthm, amsfonts, epsfig}

\usepackage[linktocpage]{hyperref}
\usepackage[usenames,dvipsnames]{color}
\usepackage{epstopdf}
\usepackage{aas_macros}
\usepackage{pifont}
\definecolor{darkred}{rgb}{0.5,0,0}
\definecolor{darkgreen}{rgb}{0,0.5,0}
\definecolor{darkblue}{rgb}{0,0,0.5}
\definecolor{prussian}{rgb}{0.0, 0.19, 0.33}
\definecolor{richelectricblue}{rgb}{0.03, 0.57, 0.82}
\definecolor{teal}{rgb}{0.0, 0.5, 0.5}
\definecolor{mediumseagreen}{rgb}{0.24, 0.7, 0.44}
\definecolor{lust}{rgb}{0.9, 0.13, 0.13}
\definecolor{ballblue}{rgb}{0.13, 0.67, 0.8}
\definecolor{darkcyan}{rgb}{0.0, 0.55, 0.55}
\definecolor{mountainmeadow}{rgb}{0.19, 0.73, 0.56}
\definecolor{palecarmine}{rgb}{0.69, 0.25, 0.21}
\definecolor{richcarmine}{rgb}{0.84, 0.0, 0.25}
\definecolor{tangelo}{rgb}{0.98, 0.3, 0.0}
\definecolor{venetian}{rgb}{0.784,0.031,0.082}
\definecolor{bdfrance}{rgb}{0.192,0.549,0.906}

\hypersetup{colorlinks=true,
            citecolor=venetian,
            linkcolor=bdfrance,
            urlcolor=lust}

\usepackage{amsmath,amssymb}
\usepackage{tensor}
\usepackage{mathtools}
\usepackage{amsbsy}
\usepackage{bm}
\usepackage{float}

\newcommand{\be}{\begin{equation}}
\newcommand{\ee}{\end{equation}}

\newcommand{\p}{\prime}

\newcommand{\nn}{\nonumber}

\newcommand{\cI}{{\cal I}}

\newcommand{\rN}{{\rm N}}

\newcommand{\bnabla}{\boldsymbol{\nabla}}
\newcommand{\bA}{\mathbf{A}}
\newcommand{\bB}{\mathbf{B}}
\newcommand{\bE}{\mathbf{E}}
\newcommand{\bx}{\mathbf{x}}

\newcommand{\bv}{\mathbf{v}}

\newcommand{\bJ}{\mathbf{J}}

\begin{document}

\title{Pitfalls in applying gravitomagnetism to galactic rotation curve modelling}

\author{Kostas Glampedakis}
\email{kostas@um.es}
\affiliation{Departamento de F\'isica, Universidad de Murcia, Murcia, E-30100, Spain}
\affiliation{Theoretical Astrophysics, University of T\"ubingen, Auf der Morgenstelle 10, T\"ubingen, D-72076, Germany}

\author{David Ian Jones}
\email{D.I.Jones@soton.ac.uk}

\affiliation{Mathematical Sciences and STAG Research Centre, University of Southampton, Southampton SO17 1BJ, United Kingdom}

\begin{abstract}
The flatness of galaxy rotation curves at large radii is generally considered to be a significant piece of evidence in support of the existence of dark matter.  Several  studies have claimed that post-Newtonian corrections to the Newtonian equations of galaxy dynamics may remove (at least to some degree) the need for dark matter.  A few recent studies have examined these claims, and identified errors in their reasoning.  We add to this critique by giving what we consider to be particularly simple and transparent description of the errors made in these post-Newtonian calculations, some of which were of a rather technical nature, others more fundamental, e.g.\ the loss of the correct relativistic scaling, promoting small corrections to order unity changes.  Our work reinforces the orthodoxy that post-Newtonian effects are indeed too small to significantly alter galactic rotation curves, and will hopefully serve as a useful  guide for others, pointing out subtle errors that one might inadvertently make in such calculations.

\end{abstract}



\maketitle

\section{Introduction}

An important piece of observational evidence supporting the notion that there exists a large quantity of dark matter in the Universe comes from galaxy rotation 
curves \citep{galacticbook}.  
The argument is simple: if one takes the observed (i.e.\ baryonic) mass content of a typical galaxy and attempts to explain its 
velocity profile as a function of galactocentric radius, one finds that the associated Newtonian gravity is too weak to provide the required centripetal force.  A dark matter 
component is then normally introduced, to produce the necessary gravitational forces to give an equilibrium solution.  Alternatively, changes to the gravitational interaction 
itself can be invoked, most notably the theory of Modified Newtonian Dynamics (MOND) (a recent overview of the subject can be found in Ref.~\cite{mondbook}).

Irrespective of whether one assumes the addition of unseen mass, or the modification of Newtonian dynamics, it is generally accepted that post-Newtonian 
corrections to Newtonian dynamics, i.e.\ corrections coming from Einstein's theory of General Relativity (GR), are too small to 
be of importance in galactic modelling.  Indeed, the relevant dimensionless parameter controlling the size of post-Newtonian corrections is $(v/c)^2 \sim 10^{-6}$, 
for a typical galactic rotational velocity $v$ \citep{galacticbook}.  

Recently, however, a number of studies have suggested that GR effects could reduce or eliminate the need for unseen matter components
(or modifications of Newtonian dynamics) in the dynamics of galaxy rotation curves.  This includes the analysis of \citet{cetal_20}, who took \emph{Gaia} data on the motion of Galactic stars and attempted to fit the results using both a 
Newtonian model with dark matter, and a GR model without dark matter.  
They found that the Milky Way's rotation curve was equally well fit by both.  The model was generalised by \citet{aetal_22_new_GR_test} and 
\citet{aetal_22_towards_full_GR}, who again found that a purely GR analysis, without dark matter, could account for the observations.  
In the studies of \citet{astesiano22} (hereafter AR22) and \cite{AR_22_long_version}, the specific role of the post-Newtonian \emph{gravitomagnetic effect} 
was made clear in these ``GR without Dark Matter'' models.  Closely related to this, \citet{ruggiero22} argued that the gravitomagnetic field can 
act as an effective Newtonian density, and affect the galactic dynamics accordingly.

Some aspects of this body of work have already received critical appraisal. \citet{ciotti22} made a careful study of the effects of including the post-Newtonian 
gravitomagnetic terms on the analysis, for both infinitely thin and finite thickness discs.  Ciotti  found that the inclusion of the gravitomagnetic term produced 
only a small correction to the Newtonian result.  \citet{lasenby23} 
also re-examined the gravitomagnetic problem in detail, but for finite thickness discs, again 
confirming the high accuracy of the purely Newtonian results, and pointed out some errors in  previous analyses.  \citet{costa23} pointed out the importance of using physically meaningful reference frames when working in fully non-linear GR.

We add to this area of research by returning to the analytically tractable problem of gravitomagnetism in infinitely thin discs (i.e.\ the model considered in AR22).  
Our analysis is particularly simple and transparent, and re-enforces the conclusion that post-Newtonian effects really do seem to be negligible in galactic dynamics.  
The main source of novelty in our work is the clear identification of several issues in previous studies, adding to the critical appraisals  of \citet{ciotti22} and \citet{lasenby23}.  
Some of these issues were rather technical in nature, some rather more fundamental.  We hope our work will help future researchers avoid such pitfalls.

\section{The galactic gravitomagnetic model}
\label{sec:galacticmodel}

The gravitomagnetic approximation of GR for a stationary system comprises the two field equations
\begin{align}
\nabla^2 \Phi & =  4\pi G \rho, 
\label{poissonPhi2}
\\ \nn
\\
\nabla^2 \bA & =  \frac{8\pi G}{c^2} \mathbf{J}.
\label{poissonA2}
\end{align}
The first is the Newtonian Poisson equation for the potential $\Phi$ sourced by a density $\rho$.  
The second is the post-Newtonian equation that relates the vector potential $\bA$ to the mass current $\bJ = \rho \bv$.
With the introduction of the gravitoelectromagnetic fields  $\bE = -\bnabla \Phi $ and $\bB = \bnabla \times \bA$, 
it is easy to show that 
\be
\bnabla \cdot \bE = - 4\pi G \rho, \qquad \bnabla \times \bB = - \frac{8\pi G}{c^2} \bJ, 
\ee
together with $ \bnabla \cdot \bB = 0$ and $\bnabla \times \bE = 0$. The associated spacetime line element takes the form,
\be
ds^2 = - \left ( 1+ \frac{2\Phi}{c^2} \right ) c^2 dt^2 + \left (1- \frac{2\Phi}{c^2} \right ) d\bx \cdot d\bx + \frac{4}{c}  ( \bA \cdot  d\bx ) c  dt.
\label{metric2}
\ee
A disc galaxy is typically modelled as a pressureless fluid; in such a case the equations of motion in the gravitomagnetic approximation comprise 
a mass continuity equation $  \partial_t \rho + \bnabla \cdot \mathbf{J} = 0 $ and an equation of motion resembling the  `Lorentz force' law (essentially the 
geodesic equation)
\be
\frac{d \bv}{dt} = \partial_t \bv + ( \bv \cdot \bnabla ) \bv  = \bE  + 2 \bv \times \bB.
\label{lorentz2}
\ee
In addition to stationarity, we will assume axisymmetry, with a purely azimuthal fluid flow, in which case the mass continuity equation becomes trivial.  
It is then most convenient to work with standard cylindrical coordinates $\{ r, \varphi, z\}$. 

The above equations for $\Phi (r,z)$ and $\bA (r, z)$ can be written as
\begin{align}
& \partial^2_r \Phi +   \frac{1}{r} \partial_r \Phi    + \partial_z^2 \Phi= 4 \pi G \rho, 
\label{poissonPhi3}
\\ \nn
\\
&  \partial^2_z A_\varphi + \frac{1}{r} \partial_r \left ( r \partial_r A_\varphi \right ) -   \frac{1}{r^2} A_\varphi = \frac{8 \pi G}{c^2} \rho V,
\label{poissonA3}
\\ \nn
\\
&  \nabla^2 A_r - \frac{1}{r^2} A_r =0, \qquad  \nabla^2 A_z = 0,
\end{align}
with $\rho = \rho (r, z)$ and $\bv = ( v_r, v_\varphi, v_z ) = (0, V(r, z), 0)$. 
At the same time, Eq.~\eqref{lorentz2} with $\partial_t \bv =0$ leads to
\be
V^2 =  r \partial_r \Phi  - 2 V \partial_r ( r A_\varphi ), \qquad   \partial_z \Phi  =  2 V \partial_z A_\varphi.
\ee
A somewhat simpler system of equations can be obtained in terms of the rescaled potential $\psi \equiv r A_\varphi$,
\begin{align}
&  \partial^2_z \psi + \partial_r^2 \psi  - \frac{1}{r} \partial_r \psi  = \frac{8 \pi G}{c^2} r \rho V,
\label{psiODE1}
\\ \nn
\\
& V^2   = r  \partial_r \Phi - 2 V \partial_r \psi,
\label{lorentz3}
\\ \nn
\\
& \partial_z \Phi  =  \frac{ 2 V}{r} \partial_z \psi.
\label{psiPhiODE1}
\end{align}
These are the same equations used in AR22~\cite{astesiano22} (modulo a factor $2/c$ in the definition of $\psi$).


\section{Solving the gravitomagnetic equations: general comments} 
\label{sect:general_comments}

According to the above system of equations, the $\{ A_r, A_z\}$ sector is completely decoupled from the $\{ \Phi, \rho, \psi, V\}$ sector.  
We will therefore not consider the $\{ A_r, A_z\}$ sector any further.  This leaves use with a set of four equations in the four unknowns $\{ \Phi, \rho, \psi, V\}$.  
The four equations are the field equations for $\Phi$ and $\psi$ (Eqs.~\eqref{poissonPhi3} and \eqref{psiODE1}, respectively), and the radial and $z$-components of the equation of motion (Eqs.~\eqref{lorentz3} and \eqref{psiPhiODE1}, respectively).  Note that the field equations apply everywhere, while the equations of motion are only to be applied within the region containing matter (i.e.~where $\rho$ in non-zero).

The problem is clearly well-posed, in the sense of having the same number of equations as unknown functions.  However, there are infinitely many possible solutions, corresponding to different density profiles $\rho(r, z)$, or to different velocity profiles $V(r, z)$.   One could attempt to find a particular solution by specifying one or other of these quantities.  This then leaves one with four equations in \emph{three} unknowns, i.e.\ the problem is then overdetermined.  One could then solve three of the equations, and use the fourth as a consistency test, to verify if the prescribed density or velocity profile was allowed.  

One could even attempt to specify both $\rho(r, z)$ \emph{and} $V(r, z)$.  In this case, one would then solve equations \eqref{poissonPhi3} and \eqref{psiODE1} via a standard Green's function approach, to obtain the two potentials $\Phi(r, z)$ and $\psi(r, z)$.  The two equations of motion (Eqs.~\eqref{lorentz3} and \eqref{psiPhiODE1}) would then act as consistency tests.  Note, however, we do not know  \emph{a priori} the ``correct'' velocity field $V(r, z)$ for a given density field $\rho(r, z)$, and in general the solution generated would  \emph{not} solve the equations of motion  (Eqs.~\eqref{lorentz3} and \eqref{psiPhiODE1}).   Some sort of iterative process may be required to produce a self-consistent solution.

In practice, one can exploit the post-Newtonian expansion, so that one need only specify the density 
$\sigma(r, z)$, and use an integral expression to compute the small difference between the actual velocity profile and the purely Newtonian profile.  
We will describe this in Section \ref{sect:contribution}, once we have made further simplifications; 
specifically, see Eq.~\eqref{V_explicit}.


\section{A flat disc model}
\label{sec:flat}

In order to simplify our search for a solution to the above equations, we will consider the case of a flat `razor-thin' disc, with density profile
\be
\rho (r, z) = \sigma (r) \delta (z),
\ee
where $\sigma(r)$ is the disc's surface density. In this Section we derive expressions for the two potentials $\Phi(r, z)$ and $\psi(r, z)$ as integrals 
over their respective source terms $\sigma(r)$ and $\sigma(r) V(r)$.

A method for solving for $\Phi(r, z)$ is described in detail in Section 2.6.2  of Ref.~\cite{galacticbook}, whose methodology we follow closely here.    
Above and below the disc, where $\rho = 0$, Eq.~\eqref{poissonPhi3} can be solved with standard separation of variables, leading to the ``spectral component''
\be
\Phi_k (r, z) = e^{-k |z|} J_0 ( k r),
\label{Phik}
\ee
where $k>0$ is the associated (continuously varying) separation constant, and $J_m(x)$ is a Bessel function of the first kind.  
This solution incorporates the physical boundary condition of  a decaying potential at spatial infinity.

This potential is continuous, but its first derivative with respect to $z$ has a discontinuity at $z=0$:
\be
\lim_{z_0 \rightarrow 0} \partial_z  \Phi_k(r, z)  \Big |_{z=-z_0}^{z=z_0} =  - 2 k J_0 (k r) .
\label{step_in_Phi_k}
\ee

The discontinuity in the full potential can be found via vertical integration of~\eqref{poissonPhi3}, leading to
\be
\lim_{z_0 \rightarrow 0} \partial_z  \Phi(r, z) \Big |_{z=-z_0}^{z=z_0} =  4 \pi G \sigma(r) .
\label{dzPhiThin}
\ee
Taking the $k$-th spectral component of this and combining with Eq.~\eqref{step_in_Phi_k} then gives
\be
\sigma_k (r) =- \frac{k J_0 (k r)}{2\pi G},
\label{sigma_k-bessel}
\ee
where $\sigma_k(r)$ has the interpretation of the spectral component of the full surface density $\sigma(r)$ that sources 
$\Phi_k (r, z)$.  This agrees with Eq. (2-179b) of Ref.~\cite{galacticbook}. 

This spectral solution can be integrated to provide the full solution:
\be
\Phi (r, z) = \int_0^\infty dk\, s(k) \Phi_k (r, z)  =  \int_0^\infty dk\, s(k)  e^{-k |z|} J_0 ( k r) ,
\label{PhiGen1}
\ee 
written in terms of some (to be determined) function $s(k)$ that depends upon the actual surface density distribution $\sigma(r)$.  
Given the linear relation (Eq.~\eqref{poissonPhi3}) between $\Phi(r, z)$ and $\sigma(r)$, it follows that an equation identical in form 
to the first equality above must relate $\sigma(r)$ and $\sigma_k(r)$, containing exactly the same $s(k)$:
\be
\sigma (r) = \int_0^\infty dk\, s(k) \sigma_k (r) . 
\label{sigma-sigma_k}
\ee 
Substituting for $\sigma_k$ using Eq.~\eqref{sigma_k-bessel} gives
\be
\sigma (r) = - \frac{1}{2\pi G} \int_0^\infty dk\, s(k) k J_0 (k r ).
\ee
This is a Hankel transform between $s(k)$ and $-2 \pi G \sigma(r)$, and so can be inverted:
\be
s(k) = - 2 \pi G \int_0^\infty dr r J_0 ( k r) \sigma (r) .
\ee 
Inserting this in the last equality of~\eqref{PhiGen1} we finally obtain
\be
 \Phi (r, z) =  -2 \pi G \int_0^\infty dk\, 
 \left \{ e^{-k |z|} J_0 ( k r)   \int_0^\infty dr^\p r^\p J_0 ( k r^\p) \sigma (r^\p) \right \}.
 \label{PhiGen2}
\ee
This formula allows the direct calculation of $\Phi(r, z)$ anywhere off the equatorial plane for a given 
surface density $\sigma (r)$.

An analogous  procedure can be applied to the gravitomagnetic field equation~\eqref{psiODE1}, to give the potential $\psi(r, z)$ in terms 
of the current density $\sigma(r) V(r)$.  Separation of variables outside the disc leads to the solution,
\be
\psi_\lambda (r, z) = r e^{-\lambda |z| } J_1 ( \lambda r ), 
\label{psi_mode}
\ee 
where $\lambda >0$ is the separation constant. This solution is obviously decaying at $|z| \to \infty$ but diverges at $r \to \infty$;
nevertheless $A_\varphi = \psi/r \to 0 $ in the same limit. There is a discontinuity in the $z$-derivative at $z=0$:
\be
\lim_{z_0 \rightarrow 0} \partial_z  \psi_\lambda (r, z)  \Big |_{z=-z_0}^{z=z_0} =  - 2 r \lambda J_1 (\lambda r) .
\label{step_in_psi_k}
\ee

The discontinuity in the full potential can be found via vertical integration of~\eqref{psiODE1}, leading to
\be
\lim_{z_0 \rightarrow 0} \partial_z  \psi(r, z) \Big |_{z=-z_0}^{z=z_0} =  \frac{8 \pi G}{c^2}  r \sigma(r) V(r) .
\label{dzpsiThin}
\ee

Taking the $\lambda$-th spectral component of this and combining with Eq.~\eqref{step_in_psi_k} then gives
\be
 [r \sigma(r) V(r)]_\lambda =  -  \frac{ c^2}{4\pi G} r \lambda J_1(\lambda r) ,
\label{varpi-sigma-V_lambda-bessel}
\ee
where $[r \sigma(r) V(r)]_\lambda $ has the interpretation of the spectral component of the full $r \sigma(r) V(r)$ term that 
sources $\psi_\lambda (r, z)$. 
This spectral solution can be integrated to provide the full solution:
\be
\psi (r,z) = \int_0^\infty d\lambda \,q(\lambda)  \psi_\lambda (r, z)  = r \int_0^\infty d\lambda\, q(\lambda) e^{-\lambda |z| } J_1 ( \lambda r ) ,
\label{psi_integral}
\ee 
written in terms of some (to be determined) function $q(\lambda)$ that depends upon the actual surface current distribution $\sigma(r) V(r)$.  
Given the linear relation (Eq.~\eqref{psiODE1}) between $\psi(r, z)$ and the term $r \sigma(r) V(r)$, it follows that an equation identical 
in form to the first equality above must relate $r \sigma(r) V(r)$ and $\psi_\lambda(r)$, containing exactly the same $q(\lambda)$:
\be
r \sigma(r) V(r) = \int_0^\infty d\lambda \, q(\lambda) [r \sigma(r) V(r)]_\lambda .
\label{source-q-lambda}
\ee 
Substituting for $[r \sigma(r) V(r)]_\lambda$ using Eq.~\eqref{varpi-sigma-V_lambda-bessel} gives
\be
r \sigma(r) V(r) = - \frac{c^2}{4\pi G} \int_0^\infty d\lambda \, q(\lambda) r \lambda J_1(\lambda r) .
\ee
The factors of $r$ on each side cancel, leaving a Hankel transform between $q(\lambda)$ and $(-4 \pi G/c^2)  \sigma(r) V(r)$ which can be inverted:
 \be
q(\lambda) = - \frac{4\pi G}{c^2} \int_0^\infty dr\, r J_1 (\lambda r ) \sigma (r) V(r).
\label{vertint2}
\ee
The general solution for $\psi$ is then obtained by substituting Eq.~\eqref{vertint2} into the last equality of Eq.~\eqref{psi_integral}  to give
\be
\psi (r,z) =  - \frac{4\pi G}{c^2} r \int_0^\infty d\lambda\, \left \{  e^{-\lambda |z| } J_1 ( \lambda r )  
\int_0^\infty dr^\p\, r^\p J_1 (\lambda r^\p ) \sigma (r^\p) V(r^\p) \right \}.
\label{psisolgen}
\ee
This formula allows the direct calculation of $\psi(r, z)$ anywhere off the equatorial plane for prescribed $\{\sigma (r), V(r) \}$ functions.

Eqns. \eqref{PhiGen2} and \eqref{psisolgen} allow one to compute the fields $\Phi (r, z)$ and $\psi (r, z)$ in terms of their respective source terms 
$\sigma(r)$ and $\sigma(r) V(r)$.  As noted in Section \ref{sect:general_comments}, it is then necessary to ensure that the chosen source terms 
and the corresponding fields solve both the radial and $z$-components of the equations of motion, i.e. Eqs.~\eqref{lorentz3} and \eqref{psiPhiODE1}.  
However, in the case of our razor-thin discs, the $z$-component of the equation of motion no longer applies, as the $z$-derivatives of both $\Phi(r, z)$ 
and $\psi({r, z})$ are not defined in the disc (i.e.\ at $z=0$).  
It follows that our razor-thin disc solutions need only satisfy the radial equation of motion, Eq.~\eqref{lorentz3}.  To make this explicit, we can insert our 
solution for $\psi(r, z=0)$  into \eqref{lorentz3} and make use of the Bessel identity $x J_1^\p(x)  = x J_0(x) - J_1(x)$, to obtain:
\be
V^2  - V^2_\rN  =  \frac{8\pi G}{c^2} V  r  \int_0^\infty d\lambda\,  \left \{  \lambda  J_0 ( \lambda r )  
\int_0^\infty dr^\p\, r^\p J_1 (\lambda r^\p ) \sigma (r^\p) V(r^\p) \right \},
\label{lorentz4}
\ee
where we have also introduced the Newtonian velocity 
\be
V^2_\rN (r) \equiv  r \partial_r \Phi  \Big |_{z=0}.
\ee
Only when this equation is satisfied do we have a self-consistent solution to our full set of equations.

In the Appendix we provide a simplified, single Hankel mode version of this calculation that also allows a comparison with the
source-free approach of AR22 (described in Section \ref{sec:AR22}) and at the same time demonstrates the overdetermined character of
the gravitomagnetic system of equations (see our earlier comment in Section~\ref{sect:general_comments}).

\section{The source-free gravitomagnetic calculation}
\label{sec:AR22}

Before considering Eq.~\eqref{lorentz4} any further and providing an estimate for the velocity profile, we pause and describe in some detail 
the calculation of AR22.  Their flat disc model consists of the same equations found in Section~\ref{sec:galacticmodel}. 
Unsurprisingly, they find the same solution~\eqref{PhiGen2} for $\Phi (r, z)$. The point of deviation from the orthodox analysis 
of the preceding sections first appears in the calculation of $\psi (r, z)$. In AR22's analysis the field equation~\eqref{psiODE1} is replaced by
its amputated form \emph{without} a source term. From that equation they initially obtain the Hankel mode solution~\eqref{psi_mode} and subsequently 
the general homogeneous solution,
\be
\psi (r,z)  = r \int_0^\infty d\lambda\, q (\lambda) e^{-\lambda |z| } J_1 ( \lambda r ).
\label{as22psisol1}
\ee 
As a consequence of working with the source-free field equation~\eqref{psiODE1}, AR22's calculation does not
include the vertically integrated Eq.~\eqref{vertint2}. 
Their solutions for $\Phi$ and $\psi $ are subsequently inserted in Eq.~\eqref{psiPhiODE1}. After taking the limit $z\to 0$, 
\be
 \int_0^\infty d\lambda\, \lambda q (\lambda)  J_1 ( \lambda r ) =  -\frac{1}{V(r)}  \partial_z \Phi \Big |_{z=0}.
\ee
The inversion of this Hankel transform gives
\be
q (\lambda) = - \int_0^\infty dr^\p r^\p \frac{J_1 (\lambda r^\p )}{V(r^\p)}   \partial_z \Phi \Big |_{z=0}.
\label{as22tpsi1}
\ee
The next step of the AR22 analysis is to rewrite Eq.~\eqref{dzPhiThin} as
\be
\lim_{z_0 \to 0} \partial_z \Phi \Big |_{z=0}^{z=z_0} = 2 \pi G \sigma (r),
\label{dzPhiThinAR}
\ee
and use it in~\eqref{as22tpsi1}. The result is,
\be
q (\lambda) = - 2 \pi G  \int_0^\infty dr^\p r^\p  J_1 (\lambda r^\p ) \frac{ \sigma (r^\p )}{V(r^\p)}.
\label{as22tpsi2}
\ee
Combining this with Eq.~\eqref{as22psisol1},
\be
\psi (r,z)  =  - 2 \pi G r \int_0^\infty d\lambda\, \left \{ e^{-\lambda |z| } J_1 ( \lambda r )   \int_0^\infty dr^\p r^\p  
J_1 (\lambda r^\p ) \frac{ \sigma (r^\p )}{V(r^\p)} \right \}.
\label{as22psisol2}
\ee 
This formula of AR22 represents the general solution for $\psi (r,z)$ for prescribed $\{ \sigma (r), V(r) \}$ functions. 
For the remaining equation of motion~\eqref{lorentz3} we have,
\be
 V^2 (r) - V^2_\rN (r) =  - 2  r V(r)  \int_0^\infty d\lambda\, \lambda q (\lambda) J_0 (\lambda r )
 =  4 \pi G  r V(r)   \int_0^\infty d\lambda\, \left \{ \lambda  J_0 (\lambda r ) \int_0^\infty dr^\p r^\p  
J_1 (\lambda r^\p ) \frac{ \sigma (r^\p )}{V(r^\p)} \right \}.
\label{lorentz3AR}
\ee
One gets an eerie feeling that something is amiss when retracing the steps of the AR22 analysis: the above equations are supposed to
represent the gravitomagnetic physics of the flat disc galactic model but the relativistic scale $1/c^2$ has disappeared altogether. 
This anomaly, of course, has sneaked in through the omission of the gravitomagnetic source term $8\pi G r \bJ/c^2$.
As a result, if one wishes to obtain $V$ from Eq.~\eqref{lorentz3AR}, the `Newtonian' structure of that equation guarantees
that the solution will represent a leading-order modification to the Newtonian velocity $V_\rN$. AR22 purport to have done 
exactly this, according to the following manipulation of~\eqref{lorentz3AR}:
\begin{align}
 & V - \frac{V_\rN^2}{V} =  - 2  r \int_0^\infty d\lambda\, \lambda q (\lambda) J_0 (\lambda r )
 ~\Rightarrow~
\\  \nn
\\
& \int_0^\infty dr J_0 (k r ) \left (  V - \frac{V_\rN^2}{V}  \right )  
=  - 2  \int_0^\infty d\lambda\, \lambda q (\lambda) \frac{\delta (\lambda-k)}{k}
 = - 2 q (k)  = 4 \pi G  \int_0^\infty dr  r   J_1 ( k r  ) \frac{ \sigma}{V},
 \end{align}
 where we have used the Bessel function orthogonality property
 \be
  \int_0^\infty dr\, r J_\nu (k r) J_\nu (k^\p r) = \frac{\delta(k-k^\p)}{k}.
 \ee
 This is really an integral equation for $V(r)$, however, AR22 bypass this problem by simply equating the
 integrands of the second line, thus finding
 \be
V^2 (r ) - V^2_\rN (r) = 4 \pi G r \sigma (r) \frac{J_1 ( k r  )}{J_0 (k r) } .
\label{VsolAR2}
\ee
With the help of~\eqref{dzPhiThinAR} and some rearrangement this leads to the formula
\be
V(r,k) 
=  \sqrt{ V^2_\rN (r)  +  2 r \frac{J_1 ( k r  )}{J_0 (k r) }   \partial_z \Phi \Big |_{z=0} },
\label{VsolAR}
\ee
which represents the main result of AR22. As expected from our earlier comment, the removal of the $1/c^2$
relativistic scale has led to a Newtonian-order modification to the Newtonian velocity $V_\rN$. But this is just one
of the problems associated with the AR22 analysis. The step of equating the integrands in the above manipulation 
has transformed $V$ (a physical space quantity) into a function that lives in the Hankel $k$-space. This is clearly
unphysical and therefore the solution~\eqref{VsolAR} should be dismissed (this point is discussed in detail in the recent
paper by Lasenby et al.~\cite{lasenby23}). 

There is a \emph{third} problem with the analysis of AR22, in addition to the removal of $1/c^2$ and the unphysical manipulation
behind the result~\eqref{VsolAR}, and it has to do with the use of Eq.~\eqref{psiPhiODE1} between the vertical field derivatives evaluated
at $z=0$. As pointed out earlier in this paper, the cusp-like character of the razor-thin disc potentials at $z=0$ implies that these derivatives are not 
well-defined there. Even if we were to consider the flat model as the limiting case of a finite-thickness system, it is easy to show that both derivatives 
should vanish at $z=0$ (thus making~\eqref{psiPhiODE1} a trivial identity) as a consequence of the system's equatorial reflection symmetry. 
Either way, this means that Eq.~\eqref{as22tpsi1} should not be part of  the AR22 analysis. However, the removal of that equation unhinges the Hankel 
amplitude $q (\lambda)$ from the rest of the calculation unless the velocity profile $V(r)$ is a priori specified.


\section{Estimating the gravitomagnetic contribution to the velocity} 
\label{sect:contribution}

As we have seen, the orthodox analysis of the gravitomagnetic model has led to the following integral expression for
the rotational velocity (Eq.~\eqref{lorentz4}):
\be
V^2  - V^2_\rN =  \frac{8\pi G}{c^2} V  r  \int_0^\infty d\lambda\,  \left \{  \lambda  J_0 ( \lambda r )  
\int_0^\infty dr^\p\, r^\p J_1 (\lambda r^\p ) \sigma (r^\p) V(r^\p) \right \}.
\ee
We can reverse the order of the right-hand-side integrals to write,
\be
V^2  - V^2_\rN =  \frac{8\pi G}{c^2} V  r  \int_0^\infty dr^\p\, r^\p\sigma (r^\p) V(r^\p)  \cI_{01} (r,r^\p ),
\label{Veqfinal}
\ee
where
\be
\cI_{01} (r,r^\p ) =  \int_0^\infty d\lambda\,   \lambda  J_0 ( \lambda r )   J_1 (\lambda r^\p ).
\ee
These expressions are clearly unwieldy when one attempts to solve them exactly. Fortunately, for the purpose of this paper
it is sufficient to obtain an order of magnitude estimation for $V$. To this end we assume,
\be
V^2 \sim V^2_\rN \sim \frac{G M}{r}, \qquad \sigma \sim \frac{M}{r^2},
\ee
where $M$ is the typical mass of the system within a radius $r$. Then, 
\be
 \frac{8\pi G}{c^2} V  r  \int_0^\infty dr^\p\, r^\p\sigma (r^\p) V(r^\p)  \cI (r,r^\p ) 
 \sim G M r \cI_{01} \left (\frac{GM}{c^2 r} \right ) ,
\ee
with $ [\cI_{01}] = [L]^{-2}$, where $[L]$ denotes dimensions of length. The post-Newtonian parameter $ GM/c^2r  \sim V_\rN^2/c^2 $ is obviously $\ll 1$. 
Therefore, the necessary condition for an appreciable deviation between $V$ and $V_\rN$ would be,
\be
 r \cI_{01} \left (\frac{GM}{c^2 } \right ) \sim 1.
\ee 
Some numerical experimentation reveals
\be
\cI_{01} \sim \frac{1}{r^2} \int_0^\infty dx x J_0 (x) J_1 (x) \sim \frac{1}{r^2}.
\ee
Therefore, the above condition is not satisfied and Eq.~\eqref{Veqfinal} can be seen to lead to the approximate solution
\be
V = V_\rN \left [ 1 + {\cal O} \left ( \frac{GM}{c^2 r } \right )  \right ].
\label{V-V_N}
\ee
This result displays a negligible gravitomagnetic contribution to the rotational velocity; this unsurprising result has appeared
repeatedly in the literature, for recent examples see Refs.~\cite{ciotti22, lasenby23}.

Indeed, exploiting the fact that the difference between $V$ and $V_\rN$ is small, one can simply substitute Eq.~\eqref{V-V_N} into Eq.~\eqref{Veqfinal} 
to obtain a completely closed-form result for $V$:
\be
V = V_\rN + \frac{4\pi G}{c^2} r  \int_0^\infty dr^\p\, r^\p\sigma (r^\p) V_\rN(r^\p)  \cI_{01} (r,r^\p ) .
\label{V_explicit}
\ee
In this way, one can obtain a complete solution having only prescribed the surface density distribution $\sigma(r)$.


\section{On the `strong gravitomagnetic limit' of GR}


 The so-called  `strong gravitomagnetic limit' of GR was introduced in Ref.~\cite{AR_22_long_version} as part of a study of galaxy rotation curves 
within GR, and has been claimed to be distinct to the standard, linear-gravity, gravitomagnetic approximation used so far in this paper. 
Based on that difference, Ref.~\cite{AR_22_long_version} concludes that GR gravity could replace (at least partially) the role of dark matter
in the rotational dynamics of galaxies.  In this short section we rebuke the notion of strong gravitomagnetism (and any conclusions stemming 
from it) by showing that it is based on the use of a homogeneous field equation, therefore, suffering from the scale-free pitfall discussed in the 
previous sections. 

The stationary-axisymmetric spacetime line element associated with the strong gravitomagnetic limit is found to be (this is Eq. (B1) in 
Ref.~\cite{AR_22_long_version})  
\be
ds^2 = - \left ( 1 + \frac{2\Phi}{c^2} - \frac{a^2}{c^2 r^2} \right ) c^2 dt^2 -  2 \frac{a}{c r}  cdt  r d\varphi 
+  \left ( 1 - \frac{2\Phi}{c^2}  \right ) r^2 d\varphi^2  + e^\Psi \left ( dr^2 + dz^2 \right ), 
\label{metricSGM} 
\ee
where the potentials $a, \Psi$ represent the post-Newtonian degrees of freedom. 
According to Ref.~\cite{AR_22_long_version}, the $g_{t\varphi}$ term is a factor $c$ bigger than the corresponding gravitomagnetic term in~\eqref{metric2}
and comparable to the Newtonian terms; as shown in~\cite{AR_22_long_version}, the same term leads to a mass density comparable to the Newtonian mass
density. 

The $g_{t\varphi}$ term of~\eqref{metric2} (designated as `GM') scales as
\be
g_{t\varphi}^{\rm GM} c dt d\varphi \sim \frac{A}{c}  c dt r d\varphi \sim \left ( \frac{G M}{c^2 r } \right )^{3/2}  c dt r d\varphi.
\ee
Taking the claim of~\cite{AR_22_long_version} at face value,  the $g_{t\varphi}$ term of~\eqref{metricSGM} (designated as `SGM')  should scale as
\be
g_{t\varphi}^{\rm SGM}  c dt d\varphi \sim  \frac{a}{c r } c dt r d\varphi \sim  \frac{G M}{c^2 r} c dt r d\varphi, 
\ee
which is indeed of the same order as the Newtonian term $\sim \Phi/c^2$ (this implies that the ${\cal O} (a^2)$ term in~\eqref{metricSGM} is of 
higher post-Newtonian order and can be safely omitted). This surprising result becomes less surprising if one looks at the field equation 
for $a(r,z)$ (this is Eq.~(85) in~\cite{AR_22_long_version})
\be
\partial_r^2 a - \frac{1}{r} \partial_r a + \partial_z^2 a =0.
\ee
As in the case of the AR22 formalism of Section~\ref{sec:AR22}, the use of a source-free field equation effectively removes the correct
relativistic scale of $a$, thereby promoting it to a Newtonian-order parameter through the use of other equations.



\section{On the `effective gravitomagnetic density'}
\label{sec:density}

In Ref.~\cite{ruggiero22} it is argued that gravitomagnetic field can manifest itself as an effective Newtonian density, and it is suggested that this energy 
density may perhaps have a significant impact on galactic dynamics.   In this section we repeat the key steps of the calculation of~\cite{ruggiero22} and 
provide an estimate of the size of this effective density. 

The first step is to take the divergence of the Lorentz force law~\eqref{lorentz2} (after setting $\partial_t \bv =0$),
\be
\bnabla \cdot [  ( \bv \cdot \bnabla ) \bv  ] = - 4\pi G \rho + 2 \bnabla \cdot (  \bv \times \bB ).
\label{divlorentz1}
\ee
The last term can be rewritten as,
\be
 \bnabla \cdot (  \bv \times \bB )  
 =  \bB \cdot ( \bnabla \times \bv ) + \frac{8\pi G}{c^2} \mathbf{J}  \cdot \bv,
\ee
and Eq.~\eqref{divlorentz1} becomes,
\be
\bnabla \cdot [  ( \bv \cdot \bnabla ) \bv  ] = - 4\pi G \rho \left ( 1-  \frac{4 v^2 }{c^2} \right )  + 2  \bB \cdot ( \bnabla \times \bv ).
\ee
After dropping the $v^2/c^2$ term, the equation can be rearranged in a Newtonian form, 
\be
4 \pi G (\rho + \rho_{\rm B} ) = - \bnabla \cdot [  ( \bv \cdot \bnabla ) \bv  ],
\ee
where, in accordance with Ref.~\cite{ruggiero22}, we have defined the effective gravitomagnetic density,
\be
\rho_{\rm B} \equiv  -\frac{\bB \cdot (\bnabla \times \bv )}{2\pi G}.
\ee
Evaluating this density for the  flat disc model,
\be
\rho_{\rm B} 
=  - \frac{ \partial_r \psi}{2\pi G r^2} \left ( V + r \partial_r V \right ).
\ee
Ref.~\cite{ruggiero22} offers no quantitative estimate for this density; repeating the order-of-magnitude analysis
of  Section~\ref{sect:contribution} it is straightforward to see that
\be
\rho_{\rm B} \sim \frac{M}{r^3}  \frac{GM}{c^2 r} \cI  \ll \rho ,
\ee
where $\cI$ stands for the ${\cal O} (1)$ Bessel function integrals. According to this estimate, the contribution
of the effective gravitomagnetic density is negligible  (as expected).

\section{Concluding remarks}
\label{sec:conclusions}

In this paper, we have  pointed out specific errors in recent studies of the role of gravitomagnetism in galactic dynamics.  
These errors were of several different types.  

At the highest and most fundamental level, the correct ordering of terms with respect to the relativistic scale $1/c^2$ was lost.
This error in identifying the relativistic scaling in the problem can lead to one (incorrectly) predicting order unity corrections to the Newtonian solutions.  We have demonstrated the occurrence of this type of error both in the context of the standard gravitomagnetic approximation to GR 
as well as in the so-called strong gravitomagnetic limit.

Errors were also made in using equations that made sense for finite thickness discs, but which lose their meaning for infinitely thin ones.  Specifically, both the 
Newtonian gravitational potential sourced by the mass distribution, and the post-Newtonian potential sourced by the mass current distribution, become cusp-like 
in the thin disc limit, such that their $z$-derivatives are simply not defined in the disc itself.  This means that the $z$-component of the equation of motion of the 
matter can no longer be applied to the system.   

Finally, we (along with \citet{lasenby23}) noted  an illegal mathematical operation in which the integrands of two integrals were equated.  This is not mathematically valid:  
the equality of two definite integrals does \emph{not} imply the equality of their integrands on a point-by-point basis.

Being careful to avoid such pitfalls, we have demonstrated that, for infinitely thin discs, the gravitomagnetic corrections to the purely Newtonian solutions are, as one 
would have expected, small, having size $(v/c)^2$ relative to the Newtonian terms.  This agrees with the analyses of \citet{ciotti22} and \citet{lasenby23}, and confirms 
that one can \emph{not} use post-Newtonian gravitomagnetic corrections to explain galactic rotation curves.  Some other explanation, whether it be dark matter or a 
modification of Newtonian dynamics, is required.

We hope our analysis clarifies the (small) role of gravitomagnetism in galactic dynamics, and helps future researchers avoid the above pitfalls.


\acknowledgments

KG acknowledges support from research grant PID2020-1149GB-I00 of the Spanish Ministerio de Ciencia e Innovaci{\'o}n.   
DIJ acknowledges support from the Science and Technologies Funding Council (STFC) via grant No. ST/R00045X/1.

\appendix

\section{Single mode solution}
\label{app:onemode}

This appendix provides a single `Hankel-mode' solution to the gravitomagnetic system of equations~\eqref{poissonPhi3}-\eqref{psiPhiODE1}.
Although this is not a realistic solution in any sense, it does help to clarify some of the points highlighted in the main text. 
Our set of equations includes the solutions to the vacuum field equations,
\be
\Phi_k (r, z) = s(k) e^{-k |z|} J_0 ( k r), 
\qquad
\psi_k (r, z) = q(k) r e^{-k |z| } J_1 ( k r ),
\label{Aeq1}
\ee
as well as the vertically-integrated inhomogeneous field equations,
\be
s(k) k J_0 (k r)  = -2 \pi G \sigma_k (r), 
\qquad 
 q(k) k  J_1 (k r ) =  -  \frac{4\pi G}{c^2} \sigma _k (r) V (r).
 \label{Aeq2}
\ee
Inserting~\eqref{Aeq1} into the radial component of the Lorentz force law, Eq.~~\eqref{lorentz3},
\be
V^2 (r)  +  s(k) k r J_1 ( k r)  = - 2 q(k) V(r)  k r  J_0 ( k r ).
\label{Vmode1}
\ee
The Hankel amplitudes $s(k), q(k)$ can be eliminated with the help of~\eqref{Aeq2} to obtain,
\be
V^2 \left [1 -  \frac{8\pi G}{c^2} \sigma_k (r)  r  \frac{J_0 ( k r )}{J_1 ( k r )} \right ]  
=  2\pi G \sigma_k (r )  r \frac{ J_1 ( k r) }{J_0 (k r)}.
\ee
This equation leads to the velocity profile,
\be
V^2 \approx  2\pi G \sigma_k   r \frac{ J_1 ( x) }{J_0 (x)}  + \left ( \frac{4 \pi G}{c} \right )^2 \sigma_k^2 r^2.
\ee
According to this orthodox single-mode analysis, the gravitomagnetic correction to the Newtonian rotational profile (i.e. the first term in the
preceding equation) is negligibly small. 

In the AR22 version of the present single-mode model the second~\eqref{Aeq2} equation is supposed to be replaced by
\be
s(k)  J_0 ( k r) =  2V (r)  q(k)   J_1 ( k r ),
\label{Aeq3}
\ee
which originates from the $z$-component of the Lorentz force law, Eq.~\eqref{psiPhiODE1} (recall that this equation is not
used in the orthodox analysis because of the ill-behaved derivatives at $z=0$). 
The corresponding solution for $V$ now takes the form
\be
V^2(r) = - s(k) k r J_1 ( k r)  + 2\pi G r \sigma_k (r)   \frac{J_0 (kr)}{J_1(kr)}.
\ee
This expression bears a clear resemblance to Eq.~\eqref{VsolAR2} of the AR22 analysis but displays an inverted Bessel function ratio in the last term.  
In contrast to the AR22 formula, the present result does not rely on the illegal step of equating integrands. 

The simultaneous use of~\eqref{Aeq3} and \eqref{Aeq2} for $q(k)$ leads to an \emph{overdetermined} system of equations; this was
to be expected, see our earlier discussion in Section~\ref{sect:general_comments}.
Indeed, the combination of these equations leads to 
\be
r \sigma_k V^2 = - \frac{c^2}{8\pi G} s(k) k r J_0( k r ).
\ee
After eliminating $\sigma_k$ with the help of the first equation~\eqref{Aeq2} we arrive at the nonsensical result $V = c/2$.


\section*{References}

\bibliography{biblio}

\begin{thebibliography}{11}%
\makeatletter
\providecommand \@ifxundefined [1]{%
 \@ifx{#1\undefined}
}%
\providecommand \@ifnum [1]{%
 \ifnum #1\expandafter \@firstoftwo
 \else \expandafter \@secondoftwo
 \fi
}%
\providecommand \@ifx [1]{%
 \ifx #1\expandafter \@firstoftwo
 \else \expandafter \@secondoftwo
 \fi
}%
\providecommand \natexlab [1]{#1}%
\providecommand \enquote  [1]{``#1''}%
\providecommand \bibnamefont  [1]{#1}%
\providecommand \bibfnamefont [1]{#1}%
\providecommand \citenamefont [1]{#1}%
\providecommand \href@noop [0]{\@secondoftwo}%
\providecommand \href [0]{\begingroup \@sanitize@url \@href}%
\providecommand \@href[1]{\@@startlink{#1}\@@href}%
\providecommand \@@href[1]{\endgroup#1\@@endlink}%
\providecommand \@sanitize@url [0]{\catcode `\\12\catcode `\$12\catcode
  `\&12\catcode `\#12\catcode `\^12\catcode `\_12\catcode `\%12\relax}%
\providecommand \@@startlink[1]{}%
\providecommand \@@endlink[0]{}%
\providecommand \url  [0]{\begingroup\@sanitize@url \@url }%
\providecommand \@url [1]{\endgroup\@href {#1}{\urlprefix }}%
\providecommand \urlprefix  [0]{URL }%
\providecommand \Eprint [0]{\href }%
\providecommand \doibase [0]{http://dx.doi.org/}%
\providecommand \selectlanguage [0]{\@gobble}%
\providecommand \bibinfo  [0]{\@secondoftwo}%
\providecommand \bibfield  [0]{\@secondoftwo}%
\providecommand \translation [1]{[#1]}%
\providecommand \BibitemOpen [0]{}%
\providecommand \bibitemStop [0]{}%
\providecommand \bibitemNoStop [0]{.\EOS\space}%
\providecommand \EOS [0]{\spacefactor3000\relax}%
\providecommand \BibitemShut  [1]{\csname bibitem#1\endcsname}%
\let\auto@bib@innerbib\@empty
\bibitem [{\citenamefont {Binney}\ and\ \citenamefont
  {Tremaine}(2008)}]{galacticbook}%
  \BibitemOpen
  \bibfield  {author} {\bibinfo {author} {\bibfnamefont {J.}~\bibnamefont
  {Binney}}\ and\ \bibinfo {author} {\bibfnamefont {S.}~\bibnamefont
  {Tremaine}},\ }\href@noop {} {\emph {\bibinfo {title} {Galactic Dynamics}}},\
  \bibinfo {edition} {2nd}\ ed.,\ Princeton Series in Astrophysics\ (\bibinfo
  {publisher} {Pinceton University Press},\ \bibinfo {year} {2008})\BibitemShut
  {NoStop}%
\bibitem [{\citenamefont {Merritt}(2020)}]{mondbook}%
  \BibitemOpen
  \bibfield  {author} {\bibinfo {author} {\bibfnamefont {D.}~\bibnamefont
  {Merritt}},\ }\href {\doibase 10.1017/9781108610926} {\emph {\bibinfo {title}
  {A Philosophical Approach to MOND}}}\ (\bibinfo  {publisher} {Cambridge
  University Press},\ \bibinfo {year} {2020})\BibitemShut {NoStop}%
\bibitem [{\citenamefont {{Crosta}}\ \emph {et~al.}(2020)\citenamefont
  {{Crosta}}, \citenamefont {{Giammaria}}, \citenamefont {{Lattanzi}},\ and\
  \citenamefont {{Poggio}}}]{cetal_20}%
  \BibitemOpen
  \bibfield  {author} {\bibinfo {author} {\bibfnamefont {M.}~\bibnamefont
  {{Crosta}}}, \bibinfo {author} {\bibfnamefont {M.}~\bibnamefont
  {{Giammaria}}}, \bibinfo {author} {\bibfnamefont {M.~G.}\ \bibnamefont
  {{Lattanzi}}}, \ and\ \bibinfo {author} {\bibfnamefont {E.}~\bibnamefont
  {{Poggio}}},\ }\href {\doibase 10.1093/mnras/staa1511} {\bibfield  {journal}
  {\bibinfo  {journal} {\mnras}\ }\textbf {\bibinfo {volume} {496}},\ \bibinfo
  {pages} {2107} (\bibinfo {year} {2020})}\BibitemShut {NoStop}%
\bibitem [{\citenamefont {{Astesiano}}\ \emph
  {et~al.}(2022{\natexlab{a}})\citenamefont {{Astesiano}}, \citenamefont
  {{Cacciatori}}, \citenamefont {{Dotti}}, \citenamefont {{Haardt}},\ and\
  \citenamefont {{Re}}}]{aetal_22_new_GR_test}%
  \BibitemOpen
  \bibfield  {author} {\bibinfo {author} {\bibfnamefont {D.}~\bibnamefont
  {{Astesiano}}}, \bibinfo {author} {\bibfnamefont {S.~L.}\ \bibnamefont
  {{Cacciatori}}}, \bibinfo {author} {\bibfnamefont {M.}~\bibnamefont
  {{Dotti}}}, \bibinfo {author} {\bibfnamefont {F.}~\bibnamefont {{Haardt}}}, \
  and\ \bibinfo {author} {\bibfnamefont {F.}~\bibnamefont {{Re}}},\ }\href
  {\doibase 10.48550/arXiv.2204.05143} {\enquote {\bibinfo {title}
  {{Re-weighting dark matter in disc galaxies: a new general relativistic
  observational test}},}\ } (\bibinfo {year} {2022}{\natexlab{a}}),\ \bibinfo
  {note} {preprint arXiv:2204.05143}\BibitemShut {NoStop}%
\bibitem [{\citenamefont {{Astesiano}}\ \emph
  {et~al.}(2022{\natexlab{b}})\citenamefont {{Astesiano}}, \citenamefont
  {{Cacciatori}}, \citenamefont {{Gorini}},\ and\ \citenamefont
  {{Re}}}]{aetal_22_towards_full_GR}%
  \BibitemOpen
  \bibfield  {author} {\bibinfo {author} {\bibfnamefont {D.}~\bibnamefont
  {{Astesiano}}}, \bibinfo {author} {\bibfnamefont {S.~L.}\ \bibnamefont
  {{Cacciatori}}}, \bibinfo {author} {\bibfnamefont {V.}~\bibnamefont
  {{Gorini}}}, \ and\ \bibinfo {author} {\bibfnamefont {F.}~\bibnamefont
  {{Re}}},\ }\href {\doibase 10.1140/epjc/s10052-022-10506-7} {\bibfield
  {journal} {\bibinfo  {journal} {Eur. Phys. J. C}\ }\textbf {\bibinfo {volume}
  {82}},\ \bibinfo {eid} {554} (\bibinfo {year}
  {2022}{\natexlab{b}})}\BibitemShut {NoStop}%
\bibitem [{\citenamefont {Astesiano}\ and\ \citenamefont
  {Ruggiero}(2022)}]{astesiano22}%
  \BibitemOpen
  \bibfield  {author} {\bibinfo {author} {\bibfnamefont {D.}~\bibnamefont
  {Astesiano}}\ and\ \bibinfo {author} {\bibfnamefont {M.~L.}\ \bibnamefont
  {Ruggiero}},\ }\href {\doibase 10.1103/PhysRevD.106.L121501} {\bibfield
  {journal} {\bibinfo  {journal} {Phys. Rev. D}\ }\textbf {\bibinfo {volume}
  {106}},\ \bibinfo {pages} {L121501} (\bibinfo {year} {2022})}\BibitemShut
  {NoStop}%
\bibitem [{\citenamefont {{Astesiano}}\ and\ \citenamefont
  {{Ruggiero}}(2022)}]{AR_22_long_version}%
  \BibitemOpen
  \bibfield  {author} {\bibinfo {author} {\bibfnamefont {D.}~\bibnamefont
  {{Astesiano}}}\ and\ \bibinfo {author} {\bibfnamefont {M.~L.}\ \bibnamefont
  {{Ruggiero}}},\ }\href {\doibase 10.1103/PhysRevD.106.044061} {\bibfield
  {journal} {\bibinfo  {journal} {\prd}\ }\textbf {\bibinfo {volume} {106}},\
  \bibinfo {eid} {044061} (\bibinfo {year} {2022})}\BibitemShut {NoStop}%
\bibitem [{\citenamefont {Ruggiero}\ \emph {et~al.}(2022)\citenamefont
  {Ruggiero}, \citenamefont {Ortolan},\ and\ \citenamefont
  {Speake}}]{ruggiero22}%
  \BibitemOpen
  \bibfield  {author} {\bibinfo {author} {\bibfnamefont {M.~L.}\ \bibnamefont
  {Ruggiero}}, \bibinfo {author} {\bibfnamefont {A.}~\bibnamefont {Ortolan}}, \
  and\ \bibinfo {author} {\bibfnamefont {C.~C.}\ \bibnamefont {Speake}},\
  }\href {\doibase 10.1088/1361-6382/ac9949} {\bibfield  {journal} {\bibinfo
  {journal} {Class. Quant. Grav.}\ }\textbf {\bibinfo {volume} {39}},\ \bibinfo
  {pages} {225015} (\bibinfo {year} {2022})}\BibitemShut {NoStop}%
\bibitem [{\citenamefont {Ciotti}(2022)}]{ciotti22}%
  \BibitemOpen
  \bibfield  {author} {\bibinfo {author} {\bibfnamefont {L.}~\bibnamefont
  {Ciotti}},\ }\href {\doibase 10.3847/1538-4357/ac82b3} {\bibfield  {journal}
  {\bibinfo  {journal} {Astrophys. J}\ }\textbf {\bibinfo {volume} {936}},\
  \bibinfo {pages} {180} (\bibinfo {year} {2022})}\BibitemShut {NoStop}%
\bibitem [{\citenamefont {Lasenby}\ \emph {et~al.}(2023)\citenamefont
  {Lasenby}, \citenamefont {Hobson},\ and\ \citenamefont {Barker}}]{lasenby23}%
  \BibitemOpen
  \bibfield  {author} {\bibinfo {author} {\bibfnamefont {A.~N.}\ \bibnamefont
  {Lasenby}}, \bibinfo {author} {\bibfnamefont {M.~P.}\ \bibnamefont {Hobson}},
  \ and\ \bibinfo {author} {\bibfnamefont {W.~E.~V.}\ \bibnamefont {Barker}},\
  }\href {\doibase 10.48550/arXiv.2303.06115} {\enquote {\bibinfo {title}
  {Gravitomagnetism and galaxy rotation curves: a cautionary tale},}\ }
  (\bibinfo {year} {2023}),\ \bibinfo {note} {preprint
  arXiv:2303.06115}\BibitemShut {NoStop}%
\bibitem [{\citenamefont {{Costa}}\ \emph {et~al.}(2023)\citenamefont
  {{Costa}}, \citenamefont {{Nat{\'a}rio}}, \citenamefont {{Frutos-Alfaro}},\
  and\ \citenamefont {{Soffel}}}]{costa23}%
  \BibitemOpen
  \bibfield  {author} {\bibinfo {author} {\bibfnamefont {L.~F.~O.}\
  \bibnamefont {{Costa}}}, \bibinfo {author} {\bibfnamefont {J.}~\bibnamefont
  {{Nat{\'a}rio}}}, \bibinfo {author} {\bibfnamefont {F.}~\bibnamefont
  {{Frutos-Alfaro}}}, \ and\ \bibinfo {author} {\bibfnamefont {M.}~\bibnamefont
  {{Soffel}}},\ }\href {\doibase 10.48550/arXiv.2303.17516} {\enquote {\bibinfo
  {title} {Reference frames in general relativity and the galactic rotation
  curves},}\ } (\bibinfo {year} {2023}),\ \bibinfo {note} {preprint
  arXiv:2303.17516}\BibitemShut {NoStop}%
\end{thebibliography}%

\end{document}